# Gyrotropic metamaterials with tailored magnetization

Nazar Pyvovar[1], Sadi Ayhan[1], Carl Pfeiffer[2], Igor Anisimov[2], Ilya Vitebskiy[2], Andrey A. Chabanov[1]

[1] Department of Physics and Astronomy, University of Texas at San Antonio, San Antonio, TX, USA.
[2] Sensors Directorate, Air Force Research Laboratory, Wright-Patterson AFB, OH, USA.

**Magnetic materials are crucial in nonreciprocal electromagnetic devices, such as isolators, circulators, and nonreciprocal phase shifters[1-3]. However, their use is often limited by the need for a uniform bias magnetic field and nonuniform demagnetizing fields, resulting in the restricted aperture of free-space devices, poor temperature stability, and incompatibility with magnetic field-sensitive applications. Alternative methods have been developed to achieve nonreciprocity using active[4-6], nonlinear[7-10], and time-varying metamaterials[11-13], each with its own advantages and limitations. Here, we present a new approach based on self-biased gyrotropic metamaterials composed of magnetically hard magnets (specifically NdFeB) embedded in a magnetically soft ferrite matrix. In this configuration, the NdFeB magnets provide the magnetic bias for the ferrite matrix, which produces a nonreciprocal response. This gyrotropic metamaterial can exhibit zero net magnetization while producing strong and uniform Faraday rotation over a broad temperature range. Without bias and demagnetizing fields, the aperture of this Faraday rotator can be virtually unlimited. Using this method, we demonstrate uniform 45-degree Faraday rotation and effective isolation across the microwave X-band.**

Ferrite materials biased by an external magnetic field are the most commonly used gyrotropic media at microwave (MW) and optical frequencies. This traditional approach involves using bulky magnets, which can be a significant problem, especially in compact devices and large-aperture applications. Alternatively, permanently magnetized materials such as hard ferrites and ferromagnets with high coercivity can be utilized. These materials can produce a nonreciprocal response, such as Faraday rotation, even without an external bias field. However, high coercivity implies that the magnetic resonance, where Faraday rotation is notably pronounced, can only be achieved at higher frequency bands[14]. Furthermore, magnetized materials generate a demagnetizing field that varies depending on their shape and may be nonuniform unless they are strictly ellipsoidal. This nonuniformity in the magnetic field can considerably affect the performance of nonreciprocal devices. Another issue common to both externally biased and self-biased magnetic materials is the presence of stray magnetic fields around the nonreciprocal device. Some critical applications, such as inertial navigation[15] and quantum sensing[16,17], cannot tolerate even the slightest stray magnetic field, yet they still require a nonreciprocal component to function properly.

To address the above problems, we propose a qualitatively different approach to achieving Faraday rotation: utilizing gyrotropic metamaterials with tailored magnetization, including configurations with zero net magnetization (ZNM), without needing a bias field. Our approach involves creating composite structures from two distinct magnetic materials, which must satisfy the following conditions:

1. One material should be magnetically hard and exhibit strong anisotropy (for example, NdFeB magnets or hard ferrites), while the other should be magnetically soft and demonstrate low loss at the operating frequency (for example, yttrium iron garnet (YIG) at MW frequencies).
2. Both magnetic materials must have high enough Curie temperatures to be magnetically saturated at room temperature. This requirement ensures that the metamaterial maintains net magnetization and Faraday rotation per unit length over a broad temperature range.

In this approach, only the magnetically soft material effectively interacts with electromagnetic radiation, thus producing Faraday rotation at the operating frequency. The magnetically hard material serves a dual purpose: it provides the magnetic bias for the magnetically soft material and compensates for its magnetization.

The proposed gyrotropic metamaterials can be compared to single-phase rare-earth ferrites[18-20], which exhibit ZNM at the magnetic compensation temperature but produce significant Faraday rotation[21,22]. These ferrites contain two distinct magnetic subsystems, such as the iron and rare-earth subsystems in rare-earth garnets[20], with their magnetizations oriented in opposite directions. The iron subsystem is fully saturated at room temperature, while the rare-earth subsystem is far from saturated. Consequently, the magnetization of the iron subsystem remains constant, whereas the contribution of the rare-earth subsystem to the net magnetization varies with temperature. Therefore, a state of ZNM can be achieved, but only at a specific temperature $T_c$; above and below $T_c$, the net magnetization of the ferrite reappears. This reappearing net magnetization can cause the ferrite to develop a domain structure[23], sharply increasing absorption while reducing or eliminating Faraday rotation. Another issue with the rare-earth ferrites is that their Faraday rotation strongly depends on temperature[22,24]. In contrast, the gyrotropic metamaterials do not have the limitations of the compensated ferrites, provided they meet the specified conditions.

The gyrotropic metamaterials can be created in two ways: by incorporating a magnetically hard material into a magnetically soft matrix or by integrating both magnetic materials within a nonmagnetic matrix. In this study, we choose the first option and develop a metamaterial composite with magnetically hard NdFeB cylinders embedded in a planar magnetically soft YIG matrix. This gyrotropic metamaterial is designed to achieve a 45-degree Faraday rotation for large-aperture MW applications in the X band.

Figures 1a and 1b provide a schematic and a photograph of the Faraday rotator based on a gyrotropic composite. We utilize a YIG disc with a diameter of 100 mm and a thickness of 4 mm, along with NdFeB cylinders measuring 3.175 mm in diameter and 9.525 mm in length. The ferrite disc features a hexagonal array of 121 holes measuring 3.3 mm in diameter and spaced 7.5 mm apart. The NdFeB cylinders are inserted into these holes, achieving a filling fraction $f = 0.1626$. The ferrite disc is placed between two impedance-matching layers made of 2.54 mm-thick Rogers TMM4 material, and the NdFeB cylinders are secured in place by 1.1 mm-thick glass wafers applied to the opposite sides of the disc. The ferrite has a reported magnetization of $M_f = 1.805$ kG, whereas the NdFeB magnets have a magnetization of $M_m = 14.5$ kG. This configuration results in a net magnetization of $M = fM_m - (1-f)M_f = 0.846$ kG. The properties of the constituent materials are described in Methods.

The Ansys Maxwell software was used to simulate the magnetic field $\boldsymbol{H}_e$ in the absence of the ferrite disc and the magnetic field $\boldsymbol{H}_0$ with the ferrite disc present (see Methods). Figure 1c shows the field magnitude $H_e$ in the $z = 0$ plane, which passes through the middle of the magnets, where $H_e$ has only a z-component, $H_e = H_{ez}$. The magnetic field gradually decreases from the center to the edges of the magnet array, resulting in an average field magnitude of approximately 2.2 kOe within a 75 mm diameter area at the center (see Supplementary Materials, Fig. S1). This field strength is sufficient to magnetize the ferrite disc beyond its saturation point.

The magnetic field $H_e$ is modified in the ferrite disc to $H_0 = H_e - [N](1-f)M_f$, where $[N]$ represents the demagnetization tensor[1,2]. The internal field $H_0$ affects the interaction between the ferrite disc and the MW radiation passing through it. Figure 1d shows the field magnitude $H_0$ in the $z = 0$ plane, with a black circle indicating the boundary of the ferrite disc. The nonuniformity in $H_0$ (see Supplementary Materials, Figs. S2a and S2b) appears comparable to that of $H_e$. Additionally, the plots of the field components $H_{0z}$ and $H_{0y}$ in the z-y plane (see Supplementary Materials, Figs. S2c and S2d) reveal field variations between the magnets and a radial component of $H_0$. The nonuniformity of $H_0$ within the ferrite disc can be attributed to several factors: the considerable distance between the magnets and their small aspect ratio, the significant spacing between the holes relative to the thickness of the ferrite disc, and the net magnetization of the composite.

The MW permeability of ferrites is described by the Polder tensor[25], $[\mu]$. For the ferrite disc of Fig. 1 with radial MW fields, the permeability tensor is given by

$$[\mu] = \begin{bmatrix} \mu & i\kappa & 0 \\ -i\kappa & \mu & 0 \\ 0 & 0 & 1 \end{bmatrix},$$

with the elements

$$\mu = 1 + \frac{\omega_M(\omega_0 + i\alpha\omega)}{(\omega_0 + i\alpha\omega)^2 - \omega^2}, \qquad \kappa = \frac{\omega_M \omega}{(\omega_0 + i\alpha\omega)^2 - \omega^2},$$

where $\omega_0 = \gamma(H_0 + H_A)$, $\omega_M = \gamma(1-f)M_f$, $\gamma = 2.8$ MHz/Oe is the gyromagnetic ratio, $\alpha$ is the dissipation parameter, $\omega$ is the angular frequency, and $H_A$ is the effective anisotropy field due to the hole array in the ferrite disc. For circularly polarized waves, $[\mu]$ becomes diagonal with the effective permeabilities

$$\mu_\pm = \mu \pm \kappa = 1 + \frac{\omega_M}{\omega_0 + i\alpha\omega \mp \omega}.$$

Since the incident linearly polarized wave can be considered as the sum of two counter-rotating circularly polarized components of equal amplitudes, $E_+$ and $E_-$, the resulting transmitted wave is elliptically polarized and undergoes polarization rotation, i.e., Faraday rotation, due to the difference in permeabilities $\mu_+$ and $\mu_-$. The attenuation constants for the circularly polarized components also differ. Designating the incident and transmitted wave amplitudes by $E_0$ and $E_t$, we can define the complex transmission coefficient, $t = E_t/E_0 = \tau \exp(i\phi)$, where $\tau = |t|$. Then, the Faraday rotation is given by $\theta_{FR} = (\phi_+ - \phi_-)/2$, and the ellipticity is $\eta = |\tau_+ - \tau_-|/(\tau_+ + \tau_-)$[26]. Reversing the wave propagation direction changes the sign of $\kappa$, which changes the direction of Faraday rotation, thus producing a nonreciprocal response.

MW transmission measurements of the gyrotropic composite were conducted in the X-band (8-12 GHz) using the experimental setup described in Methods. The composite disc was positioned perpendicular to the incident linearly polarized MW beam, which had a diameter of 75 mm. Spectra of the co-polarized and cross-polarized components of the transmitted field were obtained by rotating the receiving antenna by 90 degrees. The measured linear polarization fields were subsequently transformed into circular polarization components. Figure 2 displays the circular polarization component transmission coefficients, $\tau_+$ and $\tau_-$, the Faraday rotation $\theta_{FR}$, and the ellipticity $\eta$. Consistent with the design goals, the gyrotropic composite exhibits a 45-degree Faraday rotation with slight ellipticity—characteristics that are commonly utilized in free-space isolators.

Additionally, we compared the measurements of the self-biased composite to those of the externally biased ferrite disc in which the NdFeB magnets were replaced with aluminum cylinders of the same dimensions. As illustrated in Fig. 2, the results showed the closest agreement when the ferrite disc was subjected to an external field of 2.2 kG.

We also tested the isolation function of the Faraday rotator. A typical free-space isolator comprises a 45-degree Faraday rotator positioned between two polarizers with their axes 45 degrees apart. In our setup, we used linearly polarized transmitting and receiving antennas aligned at 45 degrees instead of the polarizers. Figure 3 shows the unidirectional transmission of the Faraday rotator measured with this specific antenna alignment. The transmittance $S_{21}$ is greater than $-1.5$ dB when the wave propagates in the direction of self-bias. In contrast, when the wave propagation direction is reversed, the transmittance $S_{12}$ drops below $-25$ dB over a 10% bandwidth at 10 GHz.

The gyrotropic composite demonstrates adequate isolation; however, it does not qualify as a magnetic or MW metamaterial because the diameter and spacing of the NdFeB magnets are not small enough relative to their length and the operating wavelength. To address this issue, we used the Ansys Maxwell and HFSS software to simulate the magnetic and MW properties of a gyrotropic metamaterial featuring ZNM. This ZNM metamaterial consists of a hexagonal array of NdFeB micromagnets, each measuring 52.4 μm in diameter and spaced 150 μm apart, embedded in a planar YIG matrix. The micromagnets' length, 4.732 mm, matches the ferrite's thickness, which is tuned to achieve a 45-degree Faraday rotation at 10 GHz. This design results in a filling fraction $f = 0.1107$ and $M = 0$.

Figure 4a shows the z-component of the magnetic field $\boldsymbol{H}_e$ produced by the NdFeB micromagnets without the ferrite, depicted in the z-y plane intersecting the micromagnet array. The field $\boldsymbol{H}_e$ is highly uniform throughout the array, achieving a strength of $H_e = fM_m = 1.605$ kOe between the micromagnets. This field strength is sufficient to magnetize the ferrite matrix to its saturation point. As expected for ZNM metamaterials, the field $\boldsymbol{H}_0$ is zero both inside and outside the composite, except near the poles of the micromagnets (see Fig. 4b and Supplementary Materials, Fig. S3). When the ZNM metamaterial is sandwiched between two impedance-matching layers, it produces a 45-degree Faraday rotation with a further reduction in ellipticity, as demonstrated in Fig. 4c. This Faraday rotator achieves isolation exceeding 40 dB and insertion loss below 0.5 dB within the frequency range of 9 to 11.5 GHz (see Fig. 4d).

Our simulations indicate that reducing the distance between the magnets can significantly enhance the uniformity of the magnetic field within the ferrite matrix, resulting in consistent Faraday rotation throughout the entire aperture of the nonreciprocal device. Given a specific NdFeB magnet filling fraction and fixed dimensions of the ferrite matrix, this enhancement can be achieved by decreasing the diameter of both the magnets and the ferrite holes while increasing their density and aspect ratio. These adjustments are also necessary to qualify the composite as a MW metamaterial.

The proposed gyrotropic metamaterials have a unique ability to maintain ZNM across a wide temperature range. This characteristic allows them to perform exceptionally well as nonreciprocal components without requiring a bias field. Their high isolation levels and low insertion loss over large areas make them an excellent choice for applications that require high-resolution imaging or precise power transmission[27,28]. Microwave applications that employ bulky circulators or isolators, such as radar and simultaneous transmit and receive antennas, could also benefit from these planar gyrotropic metamaterials. Additionally, they are ideally suited for applications with stringent requirements regarding the magnetic field environment[29,30]. With advancements in innovative fabrication methods, such as 3D magnetic multimaterial printing[31-33] and electroplating magnetic nanowire arrays into the pores of nonmagnetic membranes[34-36], these metamaterials have the potential to serve as low-cost alternatives to rare-earth ferrites in nonreciprocal devices at MW frequencies and above.

## Methods

### Sample preparation

The YIG disc was obtained from TCI Ceramics, Inc. The manufacturer provided the following properties for the YIG at 10 GHz: a dielectric constant of 15.37, a dielectric loss tangent of $1.0 \times 10^{-4}$, a saturation magnetization of $M_\mathrm{f} = 1.805$ kG, a Curie temperature of 280°C, and a ferromagnetic resonance line width of 15.98 Oe. The grade N52 NdFeB magnets were purchased from an online store, which reports a saturation magnetization of $M_\mathrm{m} = 1.45$ kG and a Curie temperature of 310°C. The Roger TMM4 laminate has a dielectric constant of 4.7 and a loss tangent of 0.002. The glass wafer has a dielectric constant of 3.78 and a loss tangent of $1.7 \times 10^{-4}$.

After placing the glass wafer, Roger TMM4 laminate layer, and YIG disc on a magnetic steel plate, the NdFeB magnets were inserted into the holes in the YIG disc and the laminate layer. The assembly was then covered with another laminate layer and a glass wafer. Finally, the assembly was made to slide off the steel plate and secured with Scotch tape.

### MW transmission measurements

The experimental setup for MW measurements in an applied magnetic field included an axial-field magnet with a horizontal bore measuring 10.5 cm in diameter and 60 cm in length, a Rohde&Schwarz ZNA43 Vector Network Analyzer (VNA), and a pair of linearly polarized X-band horn antennas positioned 30 cm from the openings of the magnet bore and connected to the VNA via MW cables. The sample was placed at the center of the magnet bore and oriented perpendicularly to the bore axis. The interior of the bore was lined with an MW absorber, which created a 7.5 cm opening through which the MW radiation could propagate.

The measurements were normalized to the transmission coefficient of the empty magnet bore to remove the contributions from the microwave cables and the space between the sample and the horns. Additionally, a time-gating technique with a 4-ns gating window was used to suppress the effects of multiple reflections on the measured spectra in the setup.

### ANSYS Maxwell and HFSS simulations

3D magnetostatic field distributions and electromagnetic wave propagation simulations were carried out using Ansys Maxwell and HFSS finite element analysis software. The gyrotropic materials composed of NdFeB magnets and ferrite were modeled based on their properties detailed in the sample preparation section above. For the composite disc, we simulated either the entire magnet array or the entire composite. For the ZNM metamaterial, we used a unit cell with periodic boundary conditions to simulate an infinite array of micromagnets or the metamaterial. The simulations with unit cells containing one and four micromagnets produced identical spatial distributions of the magnetic fields, thus validating the results.

**Acknowledgment**

This material is based upon work funded by the Air Force Office of Scientific Research (AFOSR) under contract numbers FA9550-22-1-0290, FA9550-21-1-0145, and 24RYCOR008. It was also supported in part by the Air Force Research Laboratory Sensors Directorate through the AFOSR Summer Faculty Fellowship Program, contract numbers FA8750-15-3-6003, FA9550-15-0001, and FA9550-20-F-0005.

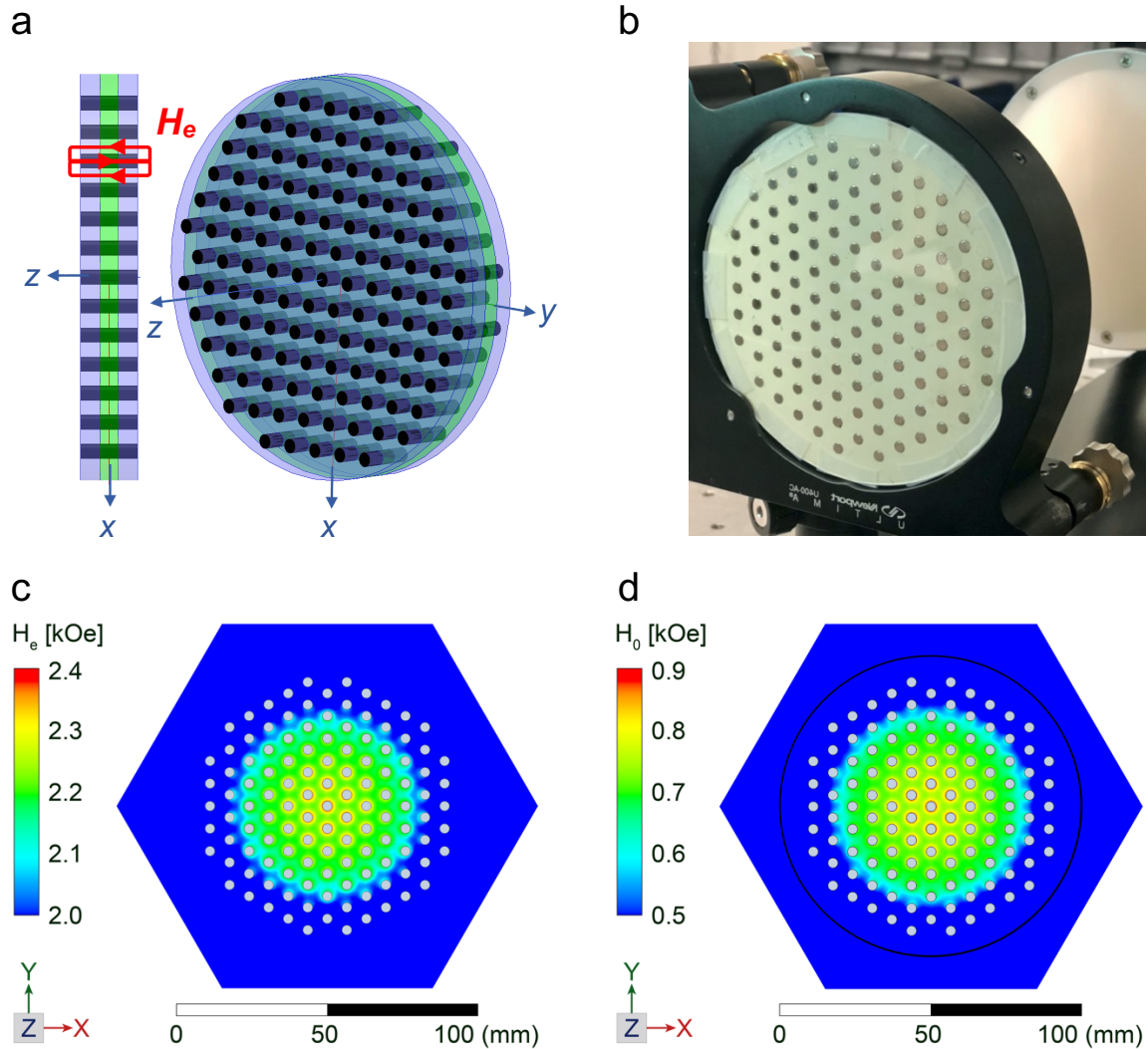

**Figure 1: Faraday rotator based on a gyrotropic composite material.** (**a**) A schematic and (**b**) a photograph of the Faraday rotator consisting of NdFeB cylinders (dark blue) incorporated into a YIG disc (green). This assembly is sandwiched between two impedance-matching layers and glass wafers. Panel (**c**) shows the magnitude of the simulated magnetic field $\boldsymbol{H}_e$ produced solely by the NdFeB magnets in the $z = 0$ plane, which passes through the middle of the magnets, where the field has only a z-component, $H_e = H_{ez}$. Panel (**d**) displays the magnitude of the simulated field $\boldsymbol{H}_0$ in the same plane when the ferrite disc is present (indicated by a black circle). The magnetic fields within the metallic NdFeB magnets (shown in gray) are not depicted, as they do not contribute to Faraday rotation.

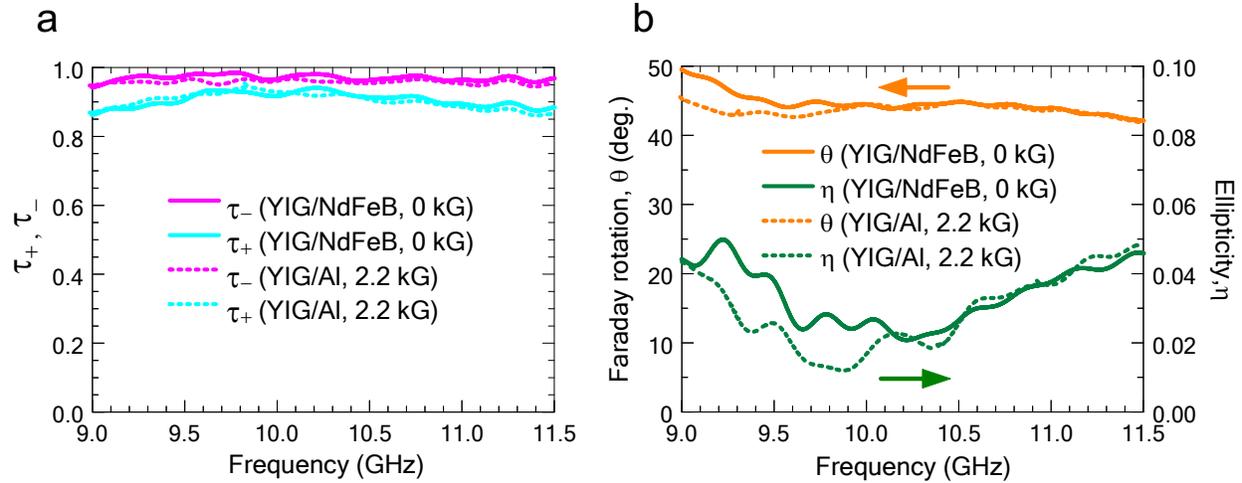

**Figure 2: Microwave transmission properties of the gyrotropic composite.** (**a**) Circularly polarized transmission coefficients and (**b**) Faraday rotation and ellipticity of the self-biased gyrotropic composite (solid lines) compared to those of the ferrite disc with aluminum cylinders replacing the NdFeB magnets under an externally applied magnetic field of 2.2 kG (dashed lines). Left-circular polarization is represented in magenta, and right-circular polarization is represented in blue. The Faraday rotation is in orange, and the ellipticity is in green.

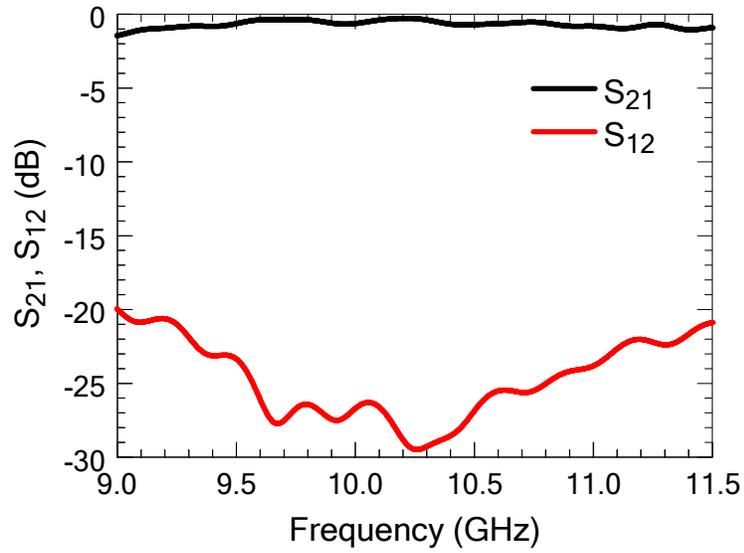

**Figure 3: Nonreciprocal transmittance of the gyrotropic composite.** Forward (black) and backward (red) microwave transmittances (S-parameters) measured in Faraday geometry with linearly polarized transmitting and receiving antennas aligned at 45 degrees.

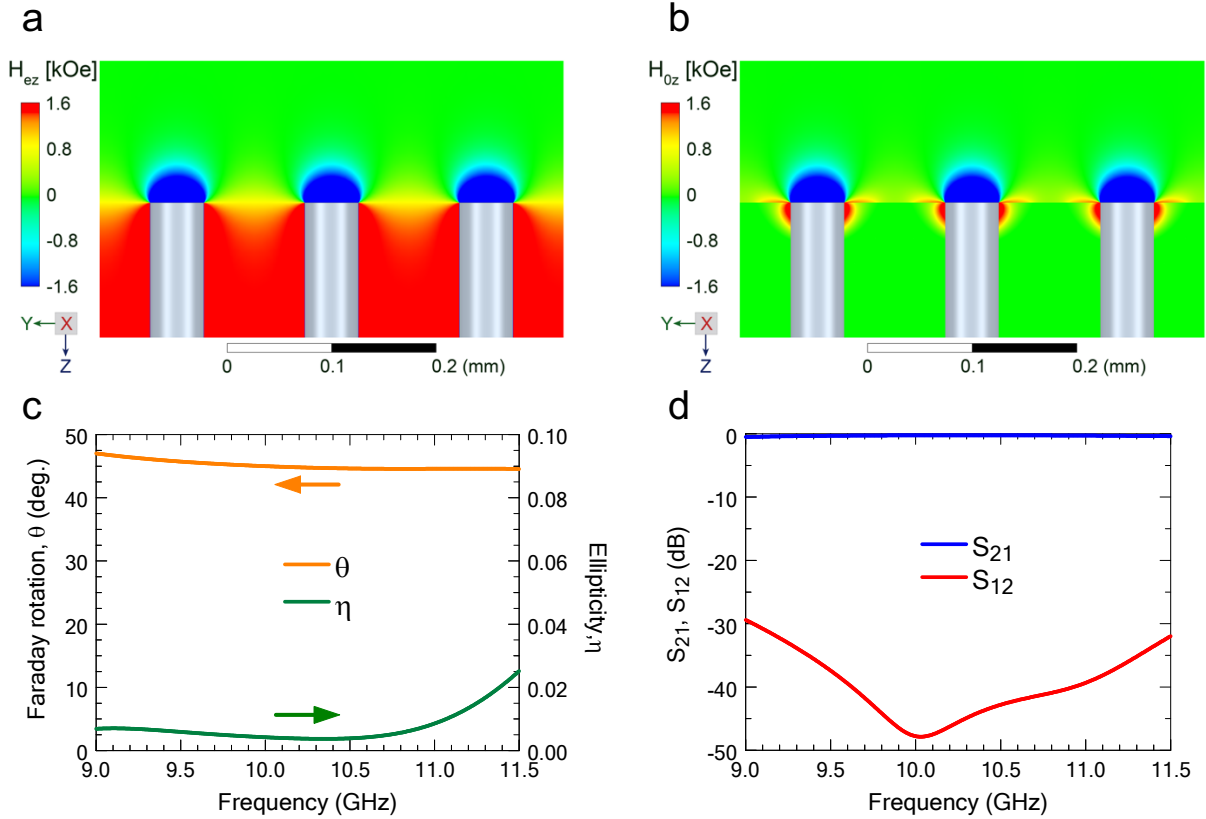

**Figure 4: Numerical simulations of a 45-degree Faraday rotator based on a gyrotropic metamaterial with zero net magnetization**. Panel (**a**) shows the z-component of the field $\boldsymbol{H}_e$ produced by a hexagonal array of NdFeB micromagnets in the absence of the ferrite matrix in the z-y plane. Panel (**b**) displays the z-component of the field $\boldsymbol{H}_0$ within and outside the ferrite matrix in the same z-y plane. Panel (**c**) illustrates the Faraday rotation and the ellipticity of the transmitted microwaves, and panel (**d**) presents the microwave isolation function.

# Supplementary Materials

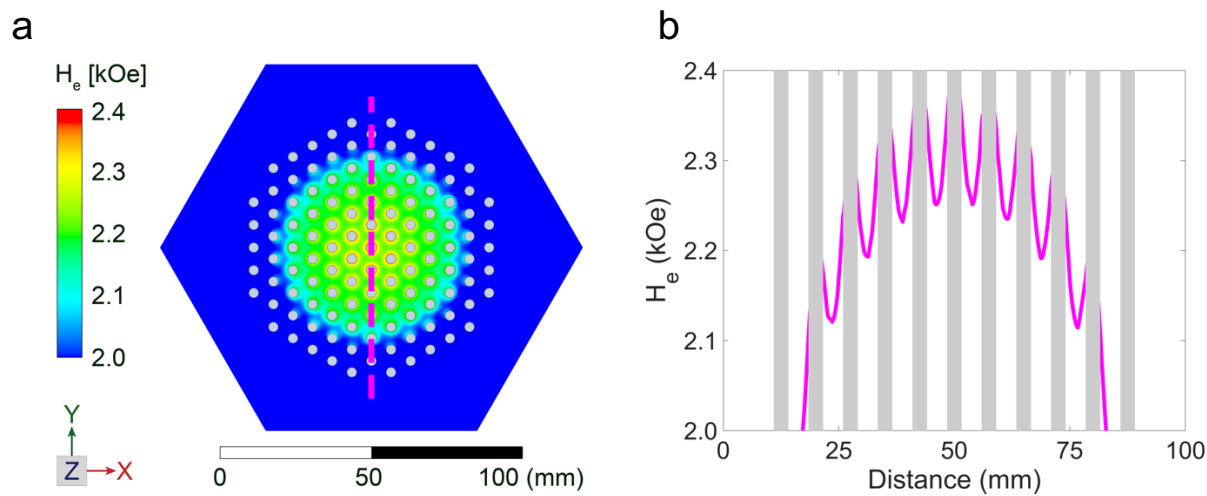

**Fig. S1.** (**a**), The spatial distribution of the simulated magnetic field $\boldsymbol{H}_e$ magnitude produced solely by the NdFeB magnets in the $z = 0$ plane that passes through their center, where the field has only a z-component, $H_e = H_{ez}$. (**b**), The graph of $H_e$ along the dashed line in panel (a) illustrates a gradual decrease of $H_e$ from the center to the edges of the magnet array while showing variations in the field between the magnets. The average field magnitude is approximately 2.2 kG within a 75 mm diameter area at the center. The magnetic field within the metallic NdFeB magnets (shown in gray) is not depicted, as they do not contribute to Faraday rotation.

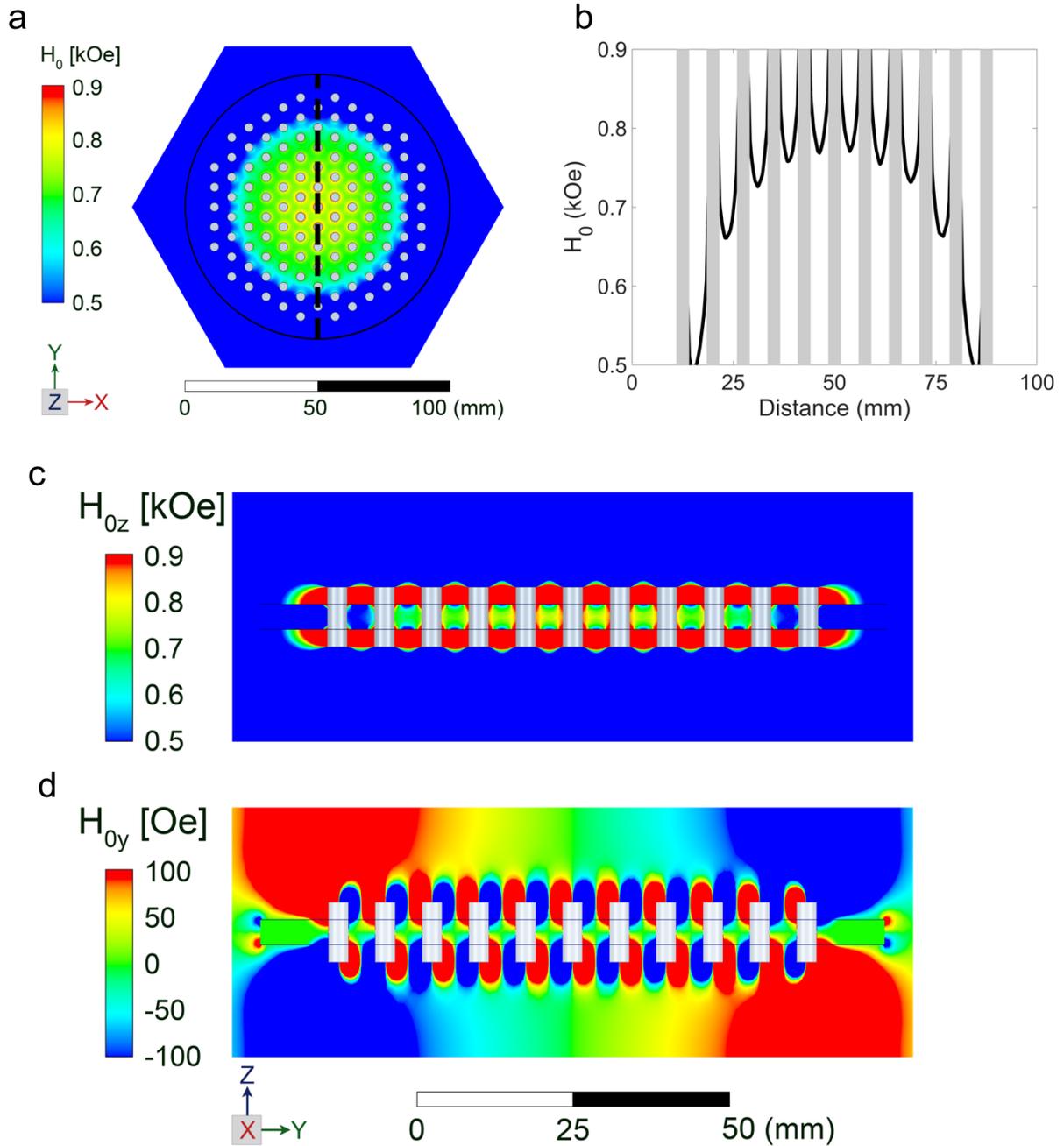

**Fig. S2.** (**a**) The magnitude of the simulated magnetic field $\boldsymbol{H}_0$ in the $z = 0$ plane passing through the middle of the ferrite disc (indicated by a black circle). In this plane, the field has only a z-component, $H_0 = H_{0z}$, due to the system's symmetry. (**b**) The field magnitude $H_0$ along the dashed line in panel (a) shows that it gradually decreases from the center to the edge of the ferrite disc, while showing variations in the field between the holes for the magnets. Plots of the (**c**) z-component and (**d**) y-component of the field $\boldsymbol{H}_0$ in the z-y plane illustrate the nonuniformity and a radial component of $\boldsymbol{H}_0$. The magnetic field within the metallic NdFeB magnets (shown in gray) is not depicted, as they do not contribute to Faraday rotation.

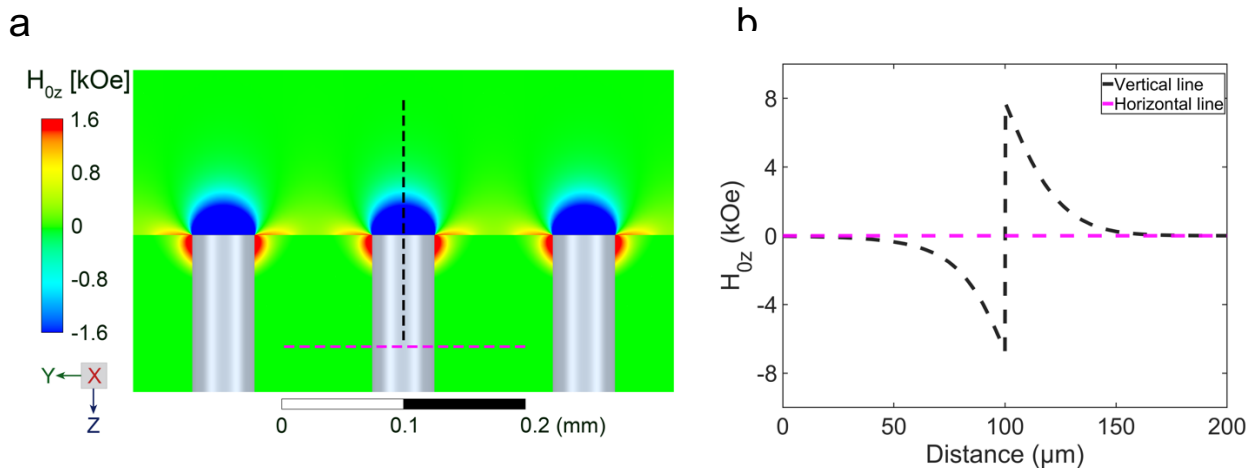

**Fig. S3. (a)** Spatial distribution of the z-component of the simulated magnetic field $\boldsymbol{H}_0$ in a gyrotropic metamaterial with zero net magnetization, which consists of NdFeB micromagnets embedded in a planar YIG matrix. The radial components of the field are zero. **(b)** The graphs of $H_{0z}$ along the dashed lines in panel (a) show that $\boldsymbol{H}_0$ rapidly approaches zero both inside (magenta line) and outside (black line) the metamaterial composite, except near the poles of the micromagnets.